\documentclass[pra,aps,twocolumn,superscriptaddress,preprint,10pt,showpacs]{revtex4-1}

\usepackage{graphicx}
\usepackage{epstopdf}
\usepackage{amsmath,amssymb,dsfont,upgreek}
\usepackage{bm}
\usepackage{natbib}
\usepackage[colorlinks=true,  linkcolor=blue, citecolor=blue, urlcolor=blue,hyperfootnotes=false]{hyperref}
\usepackage{soul}

\begin{document}
\title{Quantum theory of structured  monochromatic  light}


\author{Alexander Punnoose}
\email{apunnoose@ccny.cuny.edu}
\affiliation{Department of Physics, City College of the City University of New York, New York, NY 10031, USA}

\author{Jiufeng J. Tu}
\affiliation{Department of Physics, City College of the City University of New York, New York, NY 10031, USA}
\affiliation{Institute for Ultrafast Spectroscopy and Lasers, Department of Physics, City College of the City University of New York,  New York, NY 10031, USA}

\begin{abstract}
Applications that envisage utilizing the orbital angular momentum (OAM) at the single photon level assume that the OAM degrees of freedom that the photons inherit  from the classical wave solutions are orthogonal. To test this critical assumption, we quantize the  beam-like solutions of the vector Helmholtz equation from first principles to delineate its elementary quantum mechanical degrees of freedom.   We show that although the beam-photon operators do not in general satisfy the canonical commutation relations, implying that the photon states they create are \textit{not} orthogonal, the states are nevertheless bona fide eigenstates of the number and Hamiltonian operators. The explicit representation for the photon operators presented in this work forms a natural basis  to study light-matter interactions and quantum information processing at the single photon level.
\end{abstract}

\pacs{}
\maketitle


\section{Introduction}
Electromagnetic beams are commonly used in atomic and condensed matter physics to probe the structure of matter  by utilizing the  polarization dependence of light-matter interactions. It is known that classical optical beams can also carry a finite OAM beyond the elementary polarization state of the beam \cite{Allen_PRA,Barnett_OAM_book}. The  demonstration \cite{entanglement_Zeilinger_2001,measurement_Padgett} that the OAM defined along the direction of propagation of the beam   is carried by the individual photons therefore opens fundamentally new opportunities to study light-matter interactions at the single photon level using the transverse spatial structure of the light beams.
In solids, e.g., spectroscopic applications  using vortex light \cite{Tu_Birman_Raman,Quinteiro_LH} have been proposed to access transitions that may be weak or forbidden in conventional optical spectroscopies. Besides condensed matter applications, it has been demonstrated that this extra OAM degree of freedom of a single photon can be exploited  to encode information in photonic qubits  for quantum information processing  \cite{bobits_nature_2014}. In view of these exciting recent developments \cite{twisted_photons,book_OAM}, a general quantum theory of the traveling wave solutions of Maxwell equations that delineates the independent \textit{orthogonal} degrees of freedom  the photons inherit from the classical beams is highly desired.

To quantify the relationship between a classical light beam and the photons that make up the beam (which we call the beam-photons),  we draw on the general framework proposed by Titulaer and Glauber  (TG) \cite{glauber_PWF} to describe photons in an arbitrary basis. They introduced a new set of photon operators,   $\hat{b}_j=\sum_{\bm{k}}U_{j}(\bm{k})\hat{a}(\bm{k})$, in terms of the standard Dirac \cite{Dirac_photons,book_cohen} photon annihilation operators   $\hat{a}(\bm{k})$ in the plane-wave basis (the polarization index is suppressed for brevity); here, the information of the specific three-dimensional wave profile is arranged into the unitary matrix $U$.
When  the transformation is applied on the positive-frequency part of the electric field operator, originally written in the plane-wave basis as \cite{book_cohen}  $\bm{E}^+(\bm{r},t)=i\sum_{\bm{k}}\mathcal{E}_k \bm{e}_{\bm{k}}e^{i\bm{k}\cdot\bm{r}-i\omega_k t}\hat{a}(\bm{k})$, it reduces in the TG basis to  $\bm{E}^+(\bm{r},t)=i\sum_j \bm{\mathcal{E}}_j(\bm{r},t)\hat{b}_j$, where $\bm{\mathcal{E}}_j(\bm{r},t)=\sum_{\bm{k}}U^*_{j}(\bm{k})\mathcal{E}_k \bm{e}_{\bm{k}}e^{i\bm{k}\cdot\bm{r}-i\omega_k t}$. The key is that  since  $\omega_k=ck$ and $\bm{e}_{\bm{k}}\cdot\bm{k}=0$,   $\bm{\mathcal{E}}_j$   is guaranteed to solve Maxwell's source free equations. Conversely, given  a complete set of time-dependent classical solutions  $\bm{\mathcal{E}}_j$, the corresponding photon states in Fock space are fully specified by  $U_j$ and  the photon number in each $j$-mode. Since $U$ is unitary, the  TG operators correspond to single photon  operators that satisfy the canonical commutation relation $[\hat{b}_{j},\hat{b}^\dagger_{j'}]=\delta_{j,j'}$. In this paper, we extend this analysis to the case of monochromatic beams.

Typical problems in optics utilize  nearly monochromatic and strongly directional laser beams. For an ideal monochromatic wave, the spatial and temporal parts of the electric fields are separated as  $\bm{\mathcal{E}}_j(\bm{r},t)=e^{-i\omega t}\bm{\mathcal{E}}_j(\bm{r},\omega)$. Substituting this form into Maxwell's wave equation  gives the reduced vector Helmholtz equation \cite{book_mandel_wolf} $(\nabla^2+\omega^2/c^2)\bm{\mathcal{E}}_j(\bm{r},\omega)=0$.  Only solutions that simultaneously satisfy the auxiliary transversality condition  $\nabla\cdot\bm{\mathcal{E}}_j(\bm{r},\omega)=0$ solves the free-field Maxwell's equations.  Hence the problem of identifying the photons of a monochromatic wave reduces to quantizing the transverse solutions of the vector Helmholtz equation. A well-known family of solutions of practical relevance are the Gaussian beams, such as the Hermite-Gaussian and the Laguerre-Gaussian (LG) beams. These solutions  can be systematically obtained in a series expansion  in powers of the diffraction angle;   the lowest order terms in the series was shown by Lax \textit{et al.} \cite{Lax_paraxial} to correspond to the well-known paraxial approximation \cite{book_mandel_wolf}.
Clearly, a full quantum theory must take into account all the terms of the series. Previous works either attempted to quantize the approximate solutions in the paraxial limit \cite{Loudon_LG,LG_wunsche,LG_quant_Aiello_method}, or considered exact wave-packet (non-monochromatic) solutions of Maxwell's equations \cite{birula_RS}.

In this work,  we retain the monochromaticity condition and show rigorously to all orders beyond the paraxial limit by an explicit construction that there exists beam-photon  operators $\hat{b}_{j}^\dagger(\omega)$ and $\hat{b}_{j}(\omega)$  that create and destroy, respectively, exactly one photon with well defined quantum of  energy $\hbar\omega$. It is shown, however,  that the various orbital modes labelled by the index $j$, which the photons inherit from the classical solutions, are, unlike the polarization states,  not orthogonal in general; importantly, we show that for Gaussian beams they are not orthogonal  even in the paraxial limit, which is contrary to the conclusions arrived at in the  previous works where the orthogonality  was assumed a priori in the paraxial limit. 
In the final section we show how the paraxial limit can be recovered within our operator formalism using the LG beam as an example.

\section{Canonical quantization scheme}

The standard  plane-wave basis that used to quantize  Maxwell's equations \cite{book_cohen} does not serve well to quantize the monochromatic beam-like solutions of the vector Helmholtz equation. A more natural basis  to express the photon operators is the angular spectrum (AS) representation \cite{Aiello_AS, thesis_Visser}; it is the operator equivalent of the method of solving the Helmholtz equation for the electric and magnetic fields  by expanding them  in the plane (and evanescent) wave basis with variable amplitudes \cite{book_angular_spectrum}.

\subsection{Angular spectrum representation.}
In this section, an explicit construction of the photon operators  in frequency  space is given. The transformation of the vacuum when going from the momentum $\bm{k}$-space  to the frequency $\omega$-space is analyzed in detail.

For beam-like solutions, it is customary to label the momenta as left moving ($\bm{k}_-$) or right moving ($\bm{k}_+$) along the direction of propagation of the beam. Without loss of generality, the direction of propagation is chosen along  $\hat{\bm{z}}$, so that $\bm{k}_s=\bm{q}+s\zeta_z \hat{\bm{z}}$, where $s=\pm$ and  the momentum $\bm{q}$ is transverse  to $\hat{\bm{z}}$. Since $\zeta_z\geq 0$ in this representation,  the monochromaticity condition, $\omega_k=c|\bm{k}_s|$,  can be recast as $\zeta_z=\sqrt{\omega_k^2/c^2-q^2}$. The key observation \cite{Aiello_AS, thesis_Visser} here is that  one may define new operators  $\hat{a}_{\lambda s}(\bm{q},\omega)$ by substituting the monochromaticity condition $\omega=\omega_k$  in the plane-wave operators $\hat{a}_\lambda(\bm{k}_s)$ and integrating out $\zeta_z$;  the transformed operators $\hat{a}_{\lambda s}(\bm{q},\omega)$  can be used  to  quantize the Helmholtz equation after  suitably normalizing them to ensure that the bosonic commutation relations are satisfied. (The index $\lambda$ denotes the polarization index, which will be specified later.)

It is important to distinguish the two regions $\omega\geq cq$ and $\omega < cq$. The solutions of the  Helmholtz equation in the  region $\omega\geq cq$ are traveling (homogeneous) waves.  In the opposite region $\omega<cq$, the momentum $\zeta_z$ becomes imaginary leading to evanescent (inhomogeneous) solutions.  As shown by Carniglia and Mandel \cite{mandel_evanescent}, these latter  evanescent (inhomogeneous)  waves cannot be quantized unless additional solutions, such as  the reflected and transmitted waves around  boundaries, are included. Far from any material sources, these evanescent waves are absent and do not play a role in  any free-field theory.  Mathematically, this translates to suppressing the evanescent modes explicitly giving rise to  non-canonical commutation rules for the free-field part of the monochromatic photon operators. We show that the relevance of this constraint, while  minimal in the paraxial limit, may be significant in the case of tightly focussed beams.

The evanescent modes are suppressed explicitly by introducing new AS-operators  as follows
\begin{subequations}
\label{eqn:IAS}
\begin{eqnarray}
\hat{d}_{\lambda s}(\bm{q},\omega)&=&
\Theta(\omega-cq)\hat{a}_{\lambda s}(\bm{q},\omega)
\label{eqn:d_IAS}\\
\hat{a}_{\lambda s}(\bm{q},\omega)&=&\int_0^\infty d\zeta_z \sqrt{\mathbb{J}(q,\zeta_z)} \hat{a}_\lambda(\bm{q},s\zeta_z)\delta(\omega-\omega_k) \hspace{0.5cm}
\label{eqn:a_IAS}
\end{eqnarray}
\end{subequations}
We show that a consistent quantum theory can be constructed involving only the free-field monochromatic operators, i.e., the AS-operators $\hat{d}_{\lambda s}(\bm{q},\omega)$ and $\hat{d}^\dagger_{\lambda s}(\bm{q},\omega)$.

The operators $ \hat{a}_\lambda(\bm{q},s\zeta_z) \equiv \hat{a}_\lambda(\bm{k}_s)$  are the standard photon annihilation operators in the plane-wave basis; they satisfy the canonical relation
\begin{equation}
[\hat{a}_\lambda(\bm{q},s\zeta_z),\hat{a}_{\lambda'}^\dagger(\bm{q}',s'\zeta_z')]=\delta_{\lambda,\lambda'}\delta_{s,s'}\delta(\bm{q}-\bm{q}')\delta(\zeta_z-\zeta_z')
\label{eqn:ak_commutator}
\end{equation}
Eq.\,(\ref{eqn:a_IAS})  is an integral extension of the proposals  put forward by Visser \cite{thesis_Visser} and Aiello \textit{et al}. \cite{Aiello_AS}. The advantage of this representation is that the inverse of
 (\ref{eqn:IAS}) can be used to express the plane wave operators directly in terms of the monochromatic free-field  operators as follows
\begin{equation}
\hat{a}_\lambda(\bm{q},s\zeta_z)= \sqrt{\mathbb{J}(q,\zeta_z)} \int_0^\infty d\omega\, \hat{d}_{\lambda s}(\bm{q},\omega)\delta(\omega-\omega_k)
\label{eqn:a_kz_IAS}
\end{equation}
The  square-root factor of the Jacobian $\mathbb{J}=d\omega/d\zeta_z$  of the transformation $\omega=c\sqrt{\zeta_z^2+q^2}$ is essential to preserve the canonical commutation relations of the AS-operators (see Eq.\,(\ref{eqn:d_comm}) below); the Jacobian equals
\begin{equation}
\mathbb{J}(q,\zeta_z)=\frac{c\zeta_z}{\sqrt{\zeta_z^2+q^2}}
\label{eqn:jacobian}
\end{equation}
When $\omega\geq cq$, the argument of the  $\delta$-function in (\ref{eqn:a_IAS})  is real and can be rewritten using the identity
\begin{equation}
\delta (\omega-\omega_k)=\mathbb{J}^{-1}(q,\zeta_z)\delta (\zeta_z-\sqrt{\omega^2/c^2-q^2})
\label{eqn:delta_kz}
\end{equation}
which is well defined    on the real axis $\zeta_z=[0,\infty)$; the presence of the $\Theta$-function in (\ref{eqn:d_IAS}), defined as $\Theta(\omega-cq )=1$ when $\omega\geq cq$ and zero otherwise,  gives $\hat{d}_{\lambda s}(\bm{q},\omega)=\hat{a}_{\lambda s}(\bm{q},\omega)$.
On the other hand,   when $\omega< cq$, the actual values of $\hat{a}_{\lambda s}(\bm{q},\omega)$ are irrelevant since $\hat{d}_{\lambda s}(\bm{q},\omega)$ vanishes  by construction from the definition in (\ref{eqn:d_IAS}). The evanescent regime $\omega< cq$ is thereby explicitly suppressed.
The commutator is  obtained by substituting  the integral representation (\ref{eqn:IAS}) in the standard canonical commutator (\ref{eqn:ak_commutator}) (and simplifying using Eq.\,(\ref{eqn:delta_kz})) as
\begin{widetext}
\begin{equation}
[\hat{d}_{\lambda s}(\bm{q},\omega),\hat{d}_{\lambda' s'}^\dagger(\bm{q}',\omega')]=\delta_{\lambda,\lambda'}\delta_{s,s'}\delta(\bm{q}-\bm{q}')\delta(\omega-\omega') \Theta(\omega-c q)
\label{eqn:d_comm}
\end{equation}
\end{widetext}
Note that the  $\Theta$-factor appears explicitly above.

Finally,  the vacuum, defined in the plane wave basis as $\hat{a}_\lambda(\bm{k}_s)|0\rangle=0, \forall \bm{k}_s$, is consistently defined using  Eqs. (\ref{eqn:IAS}) and (\ref{eqn:a_kz_IAS})  as follows
\begin{subequations}
\label{eqn:vacuum}
\begin{eqnarray}
\hat{d}_{\lambda s}(\bm{q},\omega)|0\rangle &=& 0\hspace{0.5cm} \forall\, \omega, \bm{q}
\label{eqn:d_vac}\\
\hat{d}_{\lambda s}^\dagger(\bm{q},\omega)|0\rangle &=& 0\hspace{0.5cm} \omega < c q
\label{eqn:d_dagger_vac}
\end{eqnarray}
\end{subequations}

Having defined the commutators and the vacuum, we now show that $\hat{d}_{\lambda s}^\dagger(\bm{q},\omega)$ acting on an arbitrary Fock state adds a single photon with energy $\hbar \omega$ when $\omega\geq cq$. To this end, we express the Hamiltonian in the AS representation  by substituting the inverse transformation (\ref{eqn:a_kz_IAS}) in  the standard plane-wave Hamiltonian \cite{book_cohen} as follows
\begin{eqnarray}
\hat{H}&=&\sum_{\lambda, \bm{k}}\frac{\hbar\omega_k}{2}\left(\hat{a}_\lambda^\dagger(\bm{k})\hat{a}_\lambda(\bm{k})
+\hat{a}_\lambda(\bm{k})\hat{a}_\lambda^\dagger(\bm{k})\right)\\
&=&\int_0^\infty d\omega\ \frac{\hbar\omega}{2}\sum_{\lambda,s,\bm{q}} 
\left(\hat{d}_{\lambda s}^\dagger(\bm{q},\omega)\hat{d}_{\lambda s}(\bm{q},\omega)+h.c.\right)
\end{eqnarray}
Ignoring for the moment  the infinite vacuum energy, $\hat{H}$  can be written as
\begin{equation}
\hat{H}=\int_0^\infty d\omega\, \hbar\omega \hat{N}(\omega)
\label{eqn:H_IAS}
\end{equation}
where the number operator
\begin{equation}
\hat{N}(\omega)=\sum_{\lambda,s,\bm{q}}\hat{d}_{\lambda s}^\dagger(\bm{q},\omega)\hat{d}_{\lambda s}(\bm{q},\omega)
\label{eqn:n_w_IAS}
\end{equation}
corresponds to the number in frequency space; it  satisfies
\begin{equation}
[\hat{N}(\omega'),\hat{d}_{\lambda s}^\dagger(\bm{q},\omega)]=\hat{d}_{\lambda s}^\dagger(\bm{q},\omega)\delta(\omega'-\omega)\Theta(\omega-cq)
\label{eqn:comm_Nw_IAS}
\end{equation}
The total number, $\hat{N}=\int_0^\infty d\omega'\hat{N}(\omega')$, satisfies %
\begin{equation}
[\hat{N},\hat{d}_{\lambda s}^\dagger(\bm{q},\omega)]=\hat{d}_{\lambda s}^\dagger(\bm{q},\omega)\Theta(\omega-cq)
\label{eqn:comm_Na_IAS}
\end{equation}
 Eqs.\,(\ref{eqn:H_IAS})-(\ref{eqn:comm_Na_IAS}) prove that $\hat{d}_{\lambda s}^\dagger(\bm{q},\omega)$ acting on an arbitrary Fock state adds a single photon with energy $\hbar \omega$, independent of $q$, provided $\omega\geq cq$. The number states satisfy $\hat{N}|n_1,n_2,\cdots\rangle=\sum_i\Theta(\omega_i-cq_i)n_i|n_1,n_2,\cdots\rangle$; here $i=(\bm{q}_i,\omega_i,\lambda_i,s_i)$ is a collective photon label and $n_i$ is the corresponding number of photons.
The $\Theta$-factor in  (\ref{eqn:d_comm})  ensures that the norm $|| \hat{d}_{\lambda s}^\dagger(\bm{q},\omega)|n_1,n_2,\cdots\rangle ||=\Theta(\omega-cq)$ vanishes when $\omega < cq$, consistent with the definition of the vacuum in (\ref{eqn:d_dagger_vac}).

Eqs.\,(\ref{eqn:IAS})-(\ref{eqn:comm_Na_IAS})  provide the  framework for the quantization of free-field structured monochromatic beams in terms of the AS-operators $\hat{d}_{\lambda s}(\bm{q},\omega)$ and $\hat{d}_{\lambda s}^\dagger(\bm{q},\omega)$. 

\subsection{Beam-photon operators -- General theory.}
\label{b_IAS}

We now  extend the AS-operator formalism  to quantize the  free-field structured monochromatic beam-like solutions of the vector Helmholtz equation, which is our main objective for this work.  The extension to structured beams is  based on the observation \cite{thesis_Visser} that  a given family of beams can be characterized by  a unique unitary matrix $U_j(\bm{q})$  that depends only on the transverse momentum $\bm{q}$.   In the spirit of the TG proposal, we propose new beam-operators
\begin{equation}
\hat{b}_{\lambda s,j}(\omega)=\sum_{\bm{q}}U_j(\bm{q})\hat{d}_{\lambda s}(\bm{q},\omega)
\label{eqn:b}
\end{equation}
The transformation $U$ is assumed to be polarization ($\lambda$) independent. (The formalism may be extended with very little effort to include non-trivial vector beams that are formed by using non-separable combinations of the spatial and polarization modes \cite{vector_beams_OAM_book}.)

Formally, Eq.\,(\ref{eqn:b}) is inverted using the  completeness relation $\sum_jU_j^*(\bm{q})U_j(\bm{q}')=\delta(\bm{q}-\bm{q}')$, so that
\begin{equation}
\hat{d}_{\lambda s}(\bm{q},\omega)=\sum_jU^*_j(\bm{q})\hat{b}_{\lambda s,j}(\omega)
\label{eqn:b_inv}
\end{equation}
Unlike the TG operators of the  time-dependent  wave solutions of Maxwell's equations,  we show below that restricting the phase space to the homogenous solutions of the Helmholtz equation imposes non-trivial constraints on the commutation relations and the vacuum.

The first requirement for the vacuum given in Eq.\,(\ref{eqn:d_vac}) is satisfied by  (\ref{eqn:b})  by simply setting
\begin{equation}
b_{\lambda s,j}(\omega)|0\rangle=0 \hspace{0.5cm}\forall \omega
\label{eqn:b_vac}
\end{equation}
Eq.\,(\ref{eqn:b_inv}), on the other hand, leads to the following constraint  on the vacuum when the second condition (\ref{eqn:d_dagger_vac}) that annihilates the evanescent states is enforced
\begin{equation}
\sum_jU_j(\bm{q})\hat{b}_{\lambda s,j}^\dagger(\omega)|0\rangle=0\hspace{0.5cm} \omega < c q
\label{eqn:b_dagger_vac}
\end{equation}
This linear dependence is reflected in the  non-canonical form of the commutation relation found by substituting (\ref{eqn:b}) into the commutator (\ref{eqn:d_comm}), it reads as follows
\begin{equation}
[\hat{b}_{\lambda s,j}(\omega),\hat{b}_{\lambda' s',j'}^\dagger(\omega')]=\delta_{\lambda,\lambda'}\delta_{s,s'}\delta(\omega-\omega')F_{j,j'}(\omega)
\label{eqn:b_comm}
\end{equation}
where the elements %
\begin{equation}
F_{j,j'}(\omega)=\sum_{\bm{q}}U_{j}(\bm{q})U_{j'}^*(\bm{q})\Theta(\omega-cq)
\label{eqn:F}
\end{equation}
Note that $F$ is an Hermitian   projection matrix, i.e.,  $F^\dagger(\omega)=F(\omega)$ and $F^2(\omega)=F(\omega)$. It is therefore singular and non-invertible with eigenvalues  $1$ and $0$ for any finite $\omega$. The commutators therefore can not  be  brought into canonical form by simply rescaling the photon operators.
Hence, although the beam-photon states $\hat{b}_{\lambda s,j}^\dagger(\omega)|0\rangle$ are complete because $U$ is unitary, they  are not orthonormal. 
Nevertheless, we now demonstrate  that the states formed by the  action of $\hat{b}_{s,j}^\dagger(\omega)$ on $|0\rangle$ are bona fide eigenstates of the number operator $\hat{N}(\omega)$ and therefore from  (\ref{eqn:H_IAS})  have well defined energy equal to $\hbar\omega$ per photon.

To this end, we first substitute the photon operator (\ref{eqn:b_inv}) in Eq.\,(\ref{eqn:n_w_IAS}) for $N(\omega)$ and use the orthogonality relation $\sum_{\bm{q}}U_{j}(\bm{q})U_{j'}^*(\bm{q})=\delta_{j,j'}$  to get
\begin{equation}
\hat{N}(\omega)=\sum_{\lambda,s,j}\hat{b}_{\lambda s,j}^\dagger(\omega)\hat{b}_{\lambda s,j}(\omega)
\label{eqn:N_b}
\end{equation}
It follows from the commutation relation (\ref{eqn:b_comm}) that
\begin{eqnarray}
[\hat{N}(\omega'),\hat{b}_{\lambda s,j}^\dagger(\omega)]&=&\delta(\omega'-\omega)\sum_{j'}\hat{b}_{\lambda s,j'}^\dagger(\omega)F_{j',j}(\omega)\hspace{0.5cm}
\label{eqn:N_w_b_comm}\\
\protect{[}\hat{N},\hat{b}_{\lambda s,j}^\dagger(\omega)]&=&\sum_{j'}\hat{b}_{\lambda s,j'}^\dagger(\omega) F_{j',j}(\omega)
\label{eqn:N_b_comm}
\end{eqnarray}
Here, $\hat{N}$ as in (\ref{eqn:comm_Na_IAS}) is the total number operator.
Next, we decompose $F$, defined in (\ref{eqn:F}), as
\begin{subequations}
\label{eqn:F_dF}
\begin{eqnarray}
F_{j',j}(\omega)&=&\delta_{j',j}-\Delta F_{j',j}(\omega)\\
 \Delta F_{j',j}(\omega)&=&\sum_{\bm{q}}U_{j'}(\bm{q})U_{j}^*(\bm{q})\Theta(cq-\omega)
 \label{eqn:dF}
\end{eqnarray}
\end{subequations}
which simply follows by writing $\Theta(\omega-cq)=1-\Theta(cq-\omega)$.
We now show that the second term in the decomposition  in Eq.\,(\ref{eqn:b_dagger_vac}) annihilates the vacuum \cite{Alexios}
\begin{equation}
\sum_{j'}\hat{b}_{\lambda s,j'}^\dagger(\omega)\Delta F_{j',j}(\omega)|0\rangle=0
\label{eqn:F_vac}
\end{equation}
This result  is easily checked by rewriting the left hand side using Eq. (\ref{eqn:dF}) as
\begin{equation}
\sum_{\bm{q}}U_{j}^*(\bm{q})\left[\Theta(cq-\omega)\sum_{j'}U_{j'}(\bm{q})\hat{b}_{\lambda s,j'}^\dagger(\omega)|0\rangle\right]
 \end{equation}
and noting that the sum in the brackets vanishes when the condition in  Eq.\,(\ref{eqn:b_dagger_vac}) on the vacuum is applied. 
It follows from  Eqs. (\ref{eqn:N_w_b_comm})-(\ref{eqn:F_vac}) that
\begin{eqnarray}
\hat{N}(\omega')\hat{b}_{\lambda s,j}^\dagger(\omega)|0\rangle&=&\delta(\omega'-\omega)\hat{b}_{\lambda s,j}^\dagger(\omega)|0\rangle
\label{eqn:Nb_w_b}\\
\hat{N}\hat{b}_{\lambda s,j}^\dagger(\omega)|0\rangle&=&\hat{b}_{\lambda s,j}^\dagger(\omega)|0\rangle
\label{eqn:Nb_b}
\end{eqnarray}

Eqs.\,(\ref{eqn:Nb_w_b}) and (\ref{eqn:Nb_b}) combined with the Hamiltonian  (\ref{eqn:H_IAS}) allows us to conclude  that the state $\hat{b}_{\lambda s,j}^\dagger(\omega)|0\rangle$  corresponds to  \textit{exactly} one photon with energy $\hbar \omega$. Hence, all the physical photon states in Fock space can be generated by the successive action of  the $\hat{b}^\dagger$ operators on the vacuum. This is our main result in this work.

We emphasize that the photon states, which are bona fide eigenstates of the number and the Hamiltonian operators, do not in general form an orthonormal set in the mode-indices $j$. The overlap of the single photon states  is obtained  from  the commutation relation (\ref{eqn:b_comm})   as
 \begin{equation}
\langle 0|\hat{b}_{\lambda s,j}(\omega)\hat{b}_{\lambda' s',j'}^\dagger(\omega')|0\rangle=\delta_{\lambda,\lambda'}\delta_{s,s'}\delta(\omega-\omega')F_{j,j'}(\omega)
\label{eqn:b_overlap}
\end{equation}
This result has important implications for the encoding of quantum information in the transverse degrees of freedom of a single photon. Furthermore, the non-orthogonality  must be taken into account for the proper interpretation of multiphoton entanglement experiments involving structured light.

\subsection{Beam-photon operators -- Integer OAM case.}
\label{OAM}
The operator theory developed so far is now used to quantize Gaussian beams that carry a finite integer OAM.
%
%
Since  the frequency $\omega=c\sqrt{q^2+\zeta_z^2}$ is independent of the polar angle, $\theta_q$,   a beam-like solution with a finite angular momentum can be constructed by summing coherently  over the  angle weighted by the element \cite{bessel_hacyan,birula_RS} $U_l(\theta_q)=(2\pi)^{-\frac{1}{2}}e^{-il\theta_q}$. This generates a  Bessel beam  of integer order $l$ when $q$ is fixed. Note that Bessel beams are non-diffracting \cite{Durnin_bessel,Durnin_expt}   and the condition $\omega \geq c q$ is always satisfied (given fixed $q$), hence $\Delta F$ suitably defined without the $q$-sum vanishes \footnote{Blow \textit{et al}. \cite{Loudon_1D}  studied  the limiting case  of a one-dimensional system by setting  $q=0$, which is relevant for certain optical fibers. Substituting $\Theta(\omega-cq)=1$ in (\ref{eqn:F}), we get  $F_{j,j'}(\omega)=\delta_{j,j'}$, in agreement with Ref. \cite{Loudon_1D}.}.

%
%
%
%
%

Diffracting monochromatic beams with a finite OAM that are confined in the transverse direction can be constructed by integrating over the magnitude $q$  with an appropriately chosen weight function $V(q)$. To this end, we choose the  $U$ matrix to have the general form:
\begin{equation}
 U_{ml}(\bm{q})=\frac{1}{2\pi}e^{-il\theta_q}V_{ml}(q)
 \label{eqn:U_ml}
 \end{equation}
 %
 The mode indices  $m$ and $l$, collectively referred to as $j=\{ml\}$, denote the radial and azimuthal indices, respectively.
The form of $V$ can be further specialized by noting that a necessary and sufficient condition to obtain beam solutions propagating \textit{close} to the  $z$-axis \cite{book_mandel_wolf} is got by restricting the deviations in the transverse direction to values $q\ll \omega/c$. This restriction can be implemented by  introducing a single length scale, $w_0$, commonly called the beam waist.
We incorporate the waist size  by writing $V(q)$ in terms of  dimensionless variables $\tilde{V}$ and $w_0q$  as
\begin{equation}
V_{ml}(q)\equiv w_0\tilde{V}_{ml}(w_0q)
\label{eqn:V_tilde}
\end{equation}
In a typical example of a beam, e.g., the Gaussian beam,   $\tilde{V}_{ml}(w_0q)$  falls  rapidly  as $e^{-(w_0q)^2}$  when $w_0q\gg 1$. Clearly,  when $\omega \gg c/w_0$, the relevant $q$'s satisfy $q\ll \omega/c$.

To formalize these different regimes, it is convenient to define the dimensionless frequency parameter
\begin{equation}
f_\omega=\frac{c}{w_0\omega}
 \label{eqn:f}
 \end{equation}
which is  nothing but a measure of the diffraction angle. Physically, $f_\omega\ll  1$ corresponds to a weakly diffracting beam. Lax \textit{et al.} \cite{Lax_paraxial}  successfully developed a series solution  of the vector Helmholtz equation  in powers of $f_\omega$; they  showed that keeping only the lowest power in $f_\omega$, valid when $f_\omega\ll 1$, reproduces the paraxial limit. 

A similar analysis  is carried out here for the dependence of $F$ on $f_\omega$. Substituting Eqs. (\ref{eqn:U_ml}) and (\ref{eqn:V_tilde})  into Eq.\,(\ref{eqn:F}), the overlap matrix  reduces to
\begin{equation}
F_{ml,m'l'}(f_\omega)=\delta_{l,l'}\int_0^\infty \frac{\kappa d\kappa}{2\pi} \tilde{V}_{ml}^*(\kappa)\tilde{V}_{m'l}(\kappa)\Theta(1/f_\omega -\kappa)
\label{eqn:F_V_tilde}
\end{equation}
Here, $\kappa=w_0q$  and the argument of the $\Theta$-function is made dimensionless by rewriting it as: $\Theta(\omega-cq)=\Theta(w_0\omega/c-w_0q)$.  Since  $F$ is diagonal in the $l$ index, it is convenient to separate $F$  into two parts as (cf. Eq.\,(\ref{eqn:F_dF}))
\begin{subequations}
\begin{eqnarray}
F_{ml,m'l'}(f_\omega)&=&\delta_{l,l'}\left(\delta_{m,m'}-\Delta F_{ml,m'l}(f_\omega)\right)\\
\Delta F_{ml,m'l}(f_\omega)&=&\int_{1/f_\omega}^\infty\frac{\kappa d\kappa}{2\pi} \tilde{V}_{ml}^*(\kappa)\tilde{V}_{m'l}(\kappa)
\label{eqn:DF_gauss}
\end{eqnarray}
\end{subequations}
Our key observation is that in the case of a Gaussian function that falls off as $\tilde{V}(\kappa)\sim  e^{-\kappa^2}$  for $\kappa\gg 1$,   $\Delta F$ in (\ref{eqn:DF_gauss}) vanishes faster than any power of $f_\omega$ in the limit $f_\omega\ll 1$; this is readily seen  by taking the derivative $\partial_{f_\omega} \Delta F\sim |\tilde{V}(1/f_\omega)|^2\sim e^{-1/f_\omega^2}$. The same is true for all higher order derivatives as well. Thus, for beams with a Gaussian profile,  the expansion takes the form  $F_{ml,m'l'}=\delta_{l,l'}[\delta_{m,m'}+\mathcal{O}(e^{-1/f_\omega^2})]$  in the limit $f_\omega\ll 1$. 
The fact that the  asymptotic series for $F(f_\omega)$ is not a polynomial series is our main result in this section. 

We caution here that since $F$ is  singular and non-invertible for any finite $\omega$, the $\mathcal{O}(e^{-1/f_\omega^2})$ can never be ignored even for arbitrarily small $f_\omega$. Clearly, for general values of $f_\omega$ the non-orthogonality may play a significant role when the beams are tightly focussed.  This is a rigorous result which necessarily emerges when the evanescent regime is suppressed. It is our understanding that  it has always been implicitly assumed in the literature, by forcing canonical commutation relations on the beam-photon operators, that the beam-photons form an orthogonal basis  in the paraxial limit \cite{Loudon_LG,LG_wunsche,LG_quant_Aiello_method}; this assumption is not strictly valid and our conclusion, namely that the beam-photon states of a structured monochromatic free-field diffracting beam are never truly orthogonal, may be important  when studying multi-photon correlations. (The implications will be studied in future publications and will not be discussed further  in this paper.)

\section{Quantized solutions of the vector Helmholtz equation}
\label{sec:EM_quant}
The relationship between the solutions of the vector Helmholtz equation and the beam-photon operators are derived.  It is shown how the paraxial limit can be recovered within our operator formalism. 

As is well known, Maxwell's equations for the electric and magnetic fields separate into independent longitudinal  and transverse ($\perp$) components in the Coulomb gauge. In vacuum, the dynamical degrees are confined to the transverse components that satisfy the equations  $\partial_t\bm{E}_\perp= c\nabla\times \bm{B}$ and $\partial_t\bm{B}=-c\nabla\times\bm{E}_\perp$. The true  dynamical degrees of freedom are exposed by expressing the fields in terms of the vector potential $\bm{A}_\perp$ as  $\bm{E}_\perp=-c^{-1}\partial_t\bm{A}_\perp$ and $\bm{B}=\nabla\times\bm{A}_\perp$. The Coulomb gauge condition $\nabla\cdot \bm{A}_\perp=0$ reduces the degrees of freedom  to just two, corresponding to the two transverse polarization states of the photon.
They satisfy  the wave equation 
\begin{equation}
\Box \bm{A}_\perp\equiv\left(\frac{1}{c^2}\frac{\partial^2}{\partial t^2}-\nabla^2\right)\bm{A}_\perp=0
\label{eqn:A_wave}
\end{equation}

The wave solutions are quantized by raising $\bm{A}_\perp$ to the level of an operator in the standard way \cite{book_cohen}. The  positive-frequency part is expressed as
\begin{equation}
\bm{A}^{(+)}_{\perp}(\bm{r},t)=\sum_{\lambda=1,2}\int d^3\bm{k}\frac{\sqrt{c}\mathcal{A}_k}{2\pi} \hat{a}_\lambda(\bm{k})  \bm{e}_\lambda(\bm{k})e^{i\bm{k\cdot r}-i\omega_k t}
\label{eqn:A+}
\end{equation}
We deviate from standard notation by defining  $\mathcal{A}_k$  here as (without the extra  $\sqrt{c}/(2\pi)$ factor)
\begin{equation}
\mathcal{A}_k=\sqrt{\frac{c\hbar}{\omega_k}} \hspace{0.5cm}(\textrm{cgs units})
\label{eqn:Ak}
\end{equation}

In Eq. (\ref{eqn:A+}), $\hat{a}_{\lambda}(\bm{k})$ is the photon annihilation operator;  the labels $\lambda=1$ and $2$ stand for the two transverse photon polarizations.
%
%
The Coulomb gauge is enforced by demanding that the polarization vectors satisfy $\bm{k}\cdot \bm{e}_\lambda(\bm{k})=0$.  The two orthogonal unit vectors may always be chosen to satisfy the standard conventions:  $\bm{e}_1(\bm{k})\times\bm{e}_2(\bm{k})=\hat{\bm{k}}$, with $\bm{e}_1(-\bm{k})=-\bm{e}_1(\bm{k})$ and $\bm{e}_2(-\bm{k})=\bm{e}_2(\bm{k})$. An explicit representation is given as
\begin{equation}
\bm{e}_1(\bm{k})=\frac{\hat{\bm{n}}\times\bm{k}}{|\hat{\bm{n}}\times \bm{k}|} \hspace{0.25cm}\textrm{and}\hspace{0.25cm}
\bm{e}_2(\bm{k})=\hat{\bm{k}}\times\bm{e}_1(\bm{k})
\label{eqn:def_e}
\end{equation}
For now,  $\hat{\bm{n}}$ is an arbitrary fixed unit vector and $\hat{\bm{k}}$ is the unit vector in the direction of $\bm{k}$. 

%

Lax \textit{et al.}'s analysis \cite{Lax_paraxial} of the wave equation for the electric field is easily extended to the vector potential in the Coulomb gauge to show that the transversality condition prevents a plane-polarized field from being globally defined in a beam.  
For this reason, one usually works, following Davis \cite{davis_EM_beams}, in the Lorentz gauge in which $\nabla\cdot\bm{A}_\perp$ is finite. Nevertheless, there are clear advantages to working in the Coulomb gauge \cite{book_cohen}, e.g.,  although the equations in the Coulomb gauge are not manifestly covariant,  $\bm{A}_\perp$ itself is gauge invariant. 

These limitations are easily overcome by introducing the Hertz magnetic potential \cite{Hertz_AJP}  $\bm{A}_\perp=\nabla\times \bm{Z}$. The  more general Whittaker potential, involving both the $E$ and $H$-type potentials, have been realized to be an economical parameterization to analyze both the classical \cite{agrawal_whittaker} and the quantum \cite{birula_RS} properties of  beams.    For our purpose, as shown below, it is sufficient to recognize that it is the magnetic potential that best describes strongly \textit{plane-polarized} beams in the paraxial limit \cite{agrawal_whittaker}.
It is straightforward to verify by inspecting Eqs. (\ref{eqn:A+}) and (\ref{eqn:def_e})  that the $\bm{Z}_\lambda$'s  satisfying $\bm{A}^{(+)}_{\perp\lambda}=\nabla\times\bm{Z}^{(+)}_\lambda$ can be written as (up to a gradient term)
\begin{eqnarray}
\bm{Z}^{(+)}_1(\bm{r},t)&=&i\int \frac{d^3\bm{k}}{2\pi}\sqrt{c}\mathcal{A}_k\hat{a}_1(\bm{k})\frac{\hat{\bm{n}}}{|\bm{k}\times \hat{\bm{n}}|}  e^{i\bm{k\cdot r}-i\omega_k t} \label{eqn:Z1}\\
\bm{Z}^{(+)}_2(\bm{r},t)&=&i\int  \frac{d^3\bm{k}}{2\pi}\sqrt{c}\mathcal{A}_k \hat{a}_2(\bm{k}) \frac{(  \hat{\bm{k}}\times \hat{\bm{n}}) }{|\bm{k}\times \hat{\bm{n}}|}   e^{i\bm{k\cdot r}-i\omega_k t}\hspace{0.25cm}\label{eqn:Z2}
\end{eqnarray}
Note that both $\bm{Z}_1$ and $\bm{Z}_2$  satisfy the wave equation.
$\bm{Z}_1$  is special in that  its polarization does not vary with $\bm{k}$, i.e., it is globally plane-polarized along the constant vector $\hat{\bm{n}}$. Also note that the $\bm{Z}_\lambda$'s are free from the auxiliary transversality conditions; in particular,   $\nabla\cdot\bm{Z}_1\neq 0$. 

For concreteness, we fix $\hat{\bm{n}}$ to, say, $\hat{\bm{y}}$ and analyze  $\bm{e}_1$  (see (\ref{eqn:def_e})) in the paraxial limit:
\begin{equation}
\bm{e}_1(\bm{k})=\frac{\hat{\bm{y}}\times\bm{k}}{|\hat{\bm{y}}\times \bm{k}|}=\frac{(k_z,0,-q_x)}{\sqrt{k_z^2+q_x^2}}\stackrel{k_z\gg q_x}{\approx} \hat{\bm{x}}
\end{equation}
Hence,  choosing $\hat{\bm{n}}=\hat{\bm{y}}$ describes a strongly polarized $\bm{A}_{\perp 1}$  field along $\bm{e}_1\approx \hat{\bm{x}}$. Although $\bm{Z}_1$ is globally plane-polarized,  the Coulomb gauge condition $\nabla\cdot\bm{A}_\perp=0$  necessarily generates a small longitudinal component \cite{Lax_paraxial}   for $\bm{A}_\perp$  proportional to $(q_x/k_z) \hat{\bm{z}}\sim f_\omega \hat{\bm{z}}$.
For these reasons, the  potential $\bm{Z}_1$ is ideally suited to study  plane-polarized  beam solutions, which is the only case that we analyze in this work.
Without loss of generality, we choose   $\hat{\bm{n}}=\hat{\bm{y}}$ from here on. It follows from Eq.\,(\ref{eqn:Z1}) that $\bm{Z}_1$  can be expressed in terms of a scalar operator as
\begin{eqnarray}
\bm{Z}_1^{(+)}(\bm{r},t)&=&i\sum_s \hat{\mathbb{Z}}_{1s} (\bm{r},t) \hat{\bm{y}}
\label{eqn:Zs_bb}\\
\hat{\mathbb{Z}}_{1s}(\bm{r},t)&=&\int \frac{d^3\bm{k}_s}{2\pi}\frac{\sqrt{c}\mathcal{A}_k}{\sqrt{\zeta_z^2+q_x^2}}  \hat{a}_1(\bm{k}_s)  e^{i\bm{k}_s\cdot \bm{r}-i\omega_k t}
\label{eqn:Zs_bb_def}
\end{eqnarray}
%
%
where both the left and right moving ($s=\pm$) operators $\hat{\mathbb{Z}}_{1s}$  satisfy  the \textit{scalar} wave equation $\Box \hat{\mathbb{Z}}_{1s}=0$ with no additional auxiliary conditions. This reduction to the unconstrained scalar wave equation is the main advantage of the magnetic Hertz potential representation.

It is straightforward to expand $\hat{\mathbb{Z}}_{1s}(\bm{r},t)$ in  the beam-operator basis by first expressing $ \hat{a}_1(\bm{k}_s)$ in Eq. (\ref{eqn:Zs_bb_def}) in terms of the AS-operators $\hat{d}_{1 s}(\bm{q},\omega)$  (setting $\lambda=1$) using the inverse relation (\ref{eqn:a_kz_IAS}) as
\begin{widetext}
\begin{subequations}
\label{eqn:Z_AS}
\begin{eqnarray}
\hat{\mathbb{Z}}_{1s}(\bm{r},t)&=&\int_0^\infty d\omega \mathcal{A}_\omega e^{-i\omega(t-s z/c)}\int \frac{d^2\bm{q}}{2\pi}e^{i\bm{q}\cdot\bm{\rho}}
Z_{1s}(z,\omega;\bm{q})\hat{d}_{1s}(\bm{q},\omega) 
\label{app:eqn:Zs_bb_1}\\
Z_{1s}(z,\omega;\bm{q})&=&
\frac{\sqrt{c\,\mathbb{J}^{-1}(q,\omega)}}{\sqrt{\omega^2/c^2-q^2\sin^2\phi_q}}\,
e^{isz(\sqrt{\omega^2/c^2-q^2}-\omega/c)}\Theta(\omega/c-q)
\label{eqn:Zs_bb_2}
\end{eqnarray}
\end{subequations}
\end{widetext}
The notation for the cylindrical coordinates is as follows: $\bm{r}=(\bm{\rho},z)$ and the polar angle is oriented as $q_x=q\cos\phi_q$. We now express the AS-operators $\hat{d}_{1 s}(\bm{q},\omega)$ in terms of the beam-operators $\hat{b}_{1 s, j}(\omega)$ using Eq.\,(\ref{eqn:b_inv}). After a few elementary manipulations, we get

\begin{widetext}
\begin{subequations}
\label{eqn:Z_b}
\begin{eqnarray}
\hat{\mathbb{Z}}_{1s}(\bm{r},t)&=&\sum_{m,l}\int_0^\infty d\omega \mathcal{A}_\omega e^{-i\omega(t-s z/c)}\mathcal{Z}_{1s,ml}(\bm{r},\omega)\hat{b}_{1s,ml}(\omega)
\label{eqn:Zs_TG_1}\\
\mathcal{Z}_{1s,ml}(\bm{r},\omega)&=& f_\omega \int_0^{1/f_\omega}\frac{d^2\bm{\kappa}}{(2\pi)^2}
\frac{e^{i\bm{\kappa}\cdot\left(\frac{\bm{\rho}}{w_0}\right)}\exp\left[i s \left(\frac{z}{z_R}\right)\frac{1}{f_\omega^2}\left(\sqrt{1-f_\omega^2\kappa^2}-1\right)\right]}
{(1-f_\omega^2\kappa^2\sin^2\theta_\kappa)^{1/2}(1-f_\omega^2\kappa^2)^{1/4}}
 e^{il\theta_\kappa}\tilde{V}_{ml}^*(\kappa)
\label{eqn:Z_OAM}
\end{eqnarray}
\end{subequations}
\end{widetext}
The Rayleigh  length, defined here as  $z_R=w_0/f_\omega$, corresponds to the length scale over which the beam stays  well collimated along $z$. We see that the transverse and longitudinal directions of $(\bm{\rho},z)$ scale with $w_0$ and $z_R$, respectively.  The polar angle is specified as $\kappa_x=\kappa\cos\theta_\kappa$, where the dimensionless integration variable $\bm{\kappa}=w_0\bm{q}$.

To appreciate the integral solution in Eq.\,(\ref{eqn:Z_OAM}), we substitute  Eq.\,(\ref{eqn:Zs_TG_1})  in the wave equation $\Box \hat{\mathbb{Z}}_{1s}=0$ and obtain the reduced scalar Helmholtz equation \cite{Lax_paraxial} written entirely in terms of the dimensionless variables, $\tilde{\bm{\rho}}=\bm{\rho}/w_0$ and $\tilde{z}=z/z_R$, as
\begin{equation}
\left[\nabla^2_{\tilde{\bm{\rho}}}+2is\partial_{\tilde{z}}+f_\omega^2\partial_{\tilde{z}}^2\right]\mathcal{Z}_{1s,ml}(\bm{r},\omega)=0
\label{eqn:Zs_wave_equation}
\end{equation}
The function $\mathcal{Z}_{1s,ml}(\bm{r},\omega)$  in (\ref{eqn:Z_OAM})  is therefore the exact monochromatic beam solution of  (\ref{eqn:Zs_wave_equation})   for \textit{any} $f_\omega$ once the form of  $\tilde{V}_{ml}$ is specified \footnote{We note that the additional  factor $(1-f_\omega^2\kappa^2)^{1/4}$ in Eq.\,(\ref{eqn:Z_OAM})  is absent in the commonly proposed expressions for the non-paraxial extension of the classical modes (see, e.g., references\,\cite{barnett_non_paraxial,Barnett_OAM_book} for the specific case of the LG beams) --  since this factor was introduced to ensure the normalization of the photon number via the  Jacobian (cf.\,Eq.\,(\ref{eqn:jacobian})), it cannot be determined classically by solving the wave equation (\ref{eqn:Zs_wave_equation}).}.

For completeness, we end by studying  the converse problem of determining  $\tilde{V}_{ml}$ given an arbitrary  solution $\mathcal{Z}_{1s,ml}(\bm{r},\omega)$ of the wave equation, which is necessary to delineate the quantum degrees of freedom of classically generated structured OAM beams.
%
%
%
We show that the form of $\tilde{V}_{ml}$ is conveniently got from the paraxial limit of $\mathcal{Z}_{1s,ml}$  due to the simplified dependence of the Hertz potential on the parameter $f_\omega$.
To this end, we observe that the leading behavior of $\mathcal{Z}_{1s,ml}$  in (\ref{eqn:Z_OAM}) is  of  $\mathcal{O}(f_\omega)$, which can be made explicit by defining
 \begin{equation}
 \mathcal{Z}_{1s,ml}(\bm{r},\omega)\equiv f_\omega\tilde{\mathcal{Z}}_{1s,ml}(\tilde{\bm{\rho}},\tilde{z},f_\omega)
 \label{eqn:Z_1s_f}
 \end{equation}
 where $\tilde{\mathcal{Z}}_{1s,ml}$ corresponds to the integral in (\ref{eqn:Z_OAM}). Since the paraxial approximation corresponds to keeping terms \cite{Lax_paraxial}  up to  $\mathcal{O}(f_\omega)$, the paraxial limit is straightforwardly obtained by setting $f_\omega=0$ (keeping $z_R$ fixed) in $\tilde{\mathcal{Z}}_{1s,ml}$. Denoting the limit as $\tilde{\mathcal{Z}}_{1s,ml}^{(0)}(\tilde{\bm{\rho}},\tilde{z})\equiv\tilde{\mathcal{Z}}_{1s,ml}(\tilde{\bm{\rho}},\tilde{z},0)$, we directly arrive at the minimal Fresnel representation
 \begin{equation}
\tilde{\mathcal{Z}}_{1s,ml}^{(0)}(\tilde{\bm{\rho}},\tilde{z})
=\int\frac{d^2\bm{\kappa}}{(2\pi)^2}
e^{i\bm{\kappa}\cdot\tilde{\bm{\rho}}} e^{il\theta_\kappa}\tilde{V}_{ml}^*(\kappa)e^{-\frac{i}{2} s \tilde{z}\kappa^2}
\label{eqn:Zs_0}
\end{equation}
The factor $e^{-\frac{i}{2} s \tilde{z}\kappa^2}$ is nothing but the Fresnel propagator \cite{book_mandel_wolf}. Following the standard interpretation, we see that $\tilde{\mathcal{Z}}_{1s,ml}^{(0)}(\tilde{\bm{\rho}},\tilde{z})$  is got by using the Fresnel propagator to  propagate   the transverse (two-dimensional) Fourier transform, $\textrm{F.T.}[\tilde{\mathcal{Z}}_{1s,ml}^{(0)}(\tilde{\bm{\rho}},0)]$,  from the object plane located at $\tilde{z}=0$ to the image plane at $\tilde{z}$. Hence, the Fourier transform is  identified with
\begin{equation}
\label{eqn:V_def}
e^{-il\theta_\kappa}\tilde{V}_{ml}(\kappa)=\textrm{F.T.}[\tilde{\mathcal{Z}}_{1s,ml}^{(0)}(\tilde{\bm{\rho}},0)]^*
\end{equation}
This is a general result that can be used to determine $U_{ml}$ given  only the transverse profile of the magnetic Hertz potential  on the object plane.

As an example, explicit expression for the  beam-operators of the LG beam is derived below.
LG beams are defined  on the object plane by the family of eigenstates of a two-dimensional simple harmonic oscillator.  (We set $\hbar=1$, the mass to unity and the oscillator frequency as $w_0^{-2}$.) In cylindrical coordinates, the normalized eigenstates, $\Psi_{ml}(\tilde{\bm{\rho}})=e^{il\theta}\psi_m^{|l|}(\tilde{\rho})$, have well-defined angular momentum $l$ with the radial function defined as $\psi_m^{|l|}(\tilde{\rho})=C_m^{|l|}\tilde{\rho}^{|l|}e^{-\frac{1}{2}\tilde{\rho}^2}L_m^{|l|}(\tilde{\rho}^2)$.
Here, $l=0,\pm 1,\cdots$ and the radial index $m=0,1,\cdots$; the polar angle $\theta$ in real space is defined as $\tilde{\bm{\rho}}=\tilde{\rho}(\cos\theta,\sin\theta)$; the special functions $L_m^{|l|}$ are the associated Laguerre polynomials; the normalization constant equals $C_m^{|l|}=\sqrt{{m!}/{\pi(m+|l|)!}}$.
%

The Fourier transform  $\Phi_{ml}(\bm{\kappa})=\int d^2\tilde{\bm{\rho}}\, e^{i\bm{\kappa}\cdot\tilde{\bm{\rho}}}\Psi_{ml}^*(\tilde{\bm{\rho}})$ is well-known \cite{LG_fourier}, it reads as
\begin{subequations}
\label{eqn:FT}
\begin{eqnarray}
\Phi_{ml}(\bm{\kappa})&=&2\pi e^{i\varphi_{ml}} e^{-il\theta_\kappa}\phi_m^{|l|}(\kappa)
\label{eqn:phi_ml}\\
\phi_m^{|l|}(\kappa)&=&C_m^{|l|}\kappa^{|l|}e^{-\frac{1}{2}\kappa^2}L_m^{|l|}(\kappa^2)
\end{eqnarray}
\end{subequations}
where the additional phase $e^{i\varphi_{ml}}=[\textrm{sign}(l)]^le^{-i\pi m}e^{i\frac{\pi}{2}l}$.
Comparing the Fourier transform (\ref{eqn:FT}) with the definition of $\tilde{V}_{ml}$ in Eq.\,(\ref{eqn:V_def}), we get for the LG beams
\begin{equation}
\tilde{V}^\mathcal{LG}_{ml}(k)=2\pi e^{i\varphi_{ml}}\phi_m^{|l|}(\kappa)
\label{eqn:V_LG}
\end{equation}
(The normalization $\int \kappa d\kappa [\tilde{V}_{ml}^\mathcal{LG}(\kappa)]^*\tilde{V}^\mathcal{LG}_{m'l}(\kappa)=2\pi\delta_{m,m'}$.) Given $\tilde{V}^\mathcal{LG}_{ml}(k)$, the plane-polarized LG beam-photon operators take the form  (cf. Eqs. (\ref{eqn:b}), (\ref{eqn:U_ml}) and (\ref{eqn:V_tilde}))
\begin{equation}
\hat{b}^\mathcal{LG}_{1s,ml}(\omega)=w_0e^{i\varphi_{ml}}\sum_{\bm{q}}e^{-il\theta_q}\phi_m^{|l|}(w_0q) \hat{d}_{1s}(\bm{q},\omega)
\end{equation}
The AS-operators, $\hat{d}_{1s}(\bm{q},\omega)$, may be further expanded in the plane-wave basis using Eq.\,(\ref{eqn:IAS}).

To conclude this section, it is easily shown that $\tilde{V}^\mathcal{LG}_{ml}$ when substituted back into Eq.\,(\ref{eqn:Zs_0}) generates the well-known LG beam solution in the paraxial limit \cite{Lax_paraxial}. The details of this calculation can be found in Appendix \ref{app:Lax}.
We emphasize  that the form of $\hat{b}^\mathcal{LG}_{s1,ml}(\omega)$ is valid to all orders in $f_\omega$; no approximations were made at the operator level when going to the paraxial limit.

\section{Conclusion}
In this work, we have focussed  our investigation  on  understanding how the conserved quantities of a classical  monochromatic electromagnetic beam are distributed amongst the photons that make up the beam, which we call the beam-photons.   We have shown rigorously to all orders beyond the paraxial limit that there exists beam-photon operators, $\hat{b}_{j}^\dagger(\omega)$, that create single photons  with well defined energy $\hbar\omega$. The beam-photon operators of a Gaussian beam are, however,   not orthogonal  in the mode-index $j$ -- importantly, we show that this holds even in the paraxial limit. In the special case of a Gaussian beam with a finite orbital angular momentum, or OAM, the pair of mode-indices, $j=\{ml\}$, correspond to the radial index ($m$)   and the azimuthal index ($l$); the  photons in these cases are shown to be orthogonal in the $l$  index, but  non-orthogonal in the $m$ index for any value of the diffraction angle. Explicit expressions for the beam-photons are derived in  the plane-wave basis, which  can be directly applied to the study of light-matter interactions and photon correlations involving Gaussian OAM beams in the quantum domain.

\acknowledgments
The authors would like to thank A.\,Polychronakos,   R.\,Alfano, J.\,Secor and V. Menon (CCNY) and M. Hillery (Hunter College) for  insightful discussions. 
J.\,J.\,T.\ was partially supported by the National Science Foundation under grant no.\,$1464994$.

\begin{widetext}\appendix
\section{LG beams in the magnetic Hertz potential representation}
\label{app:Lax}

For completeness, we show here that the standard LG beam solutions in the paraxial limit can be recovered starting with the magnetic Hertz potential.
Given the form of $\tilde{V}^\mathcal{LG}_{ml}(\kappa)$ for the LG beams in Eq. (\ref{eqn:V_LG}), we  use the Fresnel propagator in (\ref{eqn:Zs_0}) to obtain  the beam solution $\tilde{\mathcal{Z}}_{1s,ml}^{(0)}(\tilde{\bm{\rho}},\tilde{z})$ for arbitrary $\tilde{z}$.  To this end, we first carry out the angular integration using the identity \cite{book_Watson}
\begin{equation}
I=\int_0^{2\pi}\frac{d\phi_\kappa}{2\pi}e^{il \phi_\kappa}e^{i\kappa\tilde{\rho}\cos(\phi_\kappa-\theta)}=e^{il(\theta+\frac{\pi}{2})}[\textrm{sign}(l)]^lJ_{|l|}(\kappa\tilde{\rho})
\end{equation}
Substituting $I$ into Eq. (\ref{eqn:Zs_0}) gives
\begin{equation}
\tilde{\mathcal{Z}}_{1s,ml}^{(0)}(\tilde{\bm{\rho}},\tilde{z})=
C_m^{|l|}e^{i l\theta}e^{i\pi m}
\int_0^\infty d\kappa e^{-\frac{1}{2}\kappa^2(1+i s \tilde{z})} \kappa^{|l|+1}L_m^{|l|}(\kappa^2)J_{|l|}(\tilde{\rho}\kappa)
\label{eqn:Zs_0_integral}
\end{equation}
This is a standard  integral listed in Eq. (7.421(4)) in Ref. \cite{book_GR}, it reads as
\begin{equation}
\int_0^\infty dx\, x^{|l|+1}e^{-\beta x^2}L_m^{|l|}(\alpha x^2)J_{|l|}(\gamma x)=\frac{(\beta-\alpha)^m\gamma^{|l|}}{2^{|l|+1}\beta^{m+|l|+1}}e^{-\frac{\gamma^2}{4\beta}}L_m^{|l|}\left[\frac{\alpha\gamma^2}{4\beta(\alpha-\beta)}\right]
\end{equation}
To compare with the integral in (\ref{eqn:Zs_0_integral}), we set  $\alpha=1, \beta=\frac{1}{2}(1+is\tilde{z}), \gamma=\tilde{\rho}$, and
%
$\beta-\alpha=-\frac{1}{2}(1-is\tilde{z})=-\beta^*$.
Substituting these parameters gives
\begin{subequations}
\label{eqn:L_LG}
\begin{eqnarray}
\tilde{\mathcal{Z}}^{(0)}_{1s,ml}&\equiv&w_0\mathcal{L}_{1s,ml}\\
\mathcal{L}_{1s,ml}&=&e^{i(l\theta-s\chi_m^{|l|})}\frac{C_m^{|l|}}{w(z)}\left(\frac{\rho}{w(z)}\right)^{|l|}
\exp\left[-\frac{1}{2}\left(\frac{\rho}{w(z)}\right)^2\right]L_n^{|l|}\left[\left(\frac{\rho}{w(z)}\right)^2\right]
\label{eqn:Y}\\
\chi_m^{| l |}&=&(2m+|l|+1)\arctan\left[\frac{z}{z_R}\right]-\frac{1}{2}\left(\frac{\rho}{w(z)}\right)^2\left(\frac{z}{z_R}\right)
\end{eqnarray}
\end{subequations}
%
The dimensionless variables are expanded  to  show the scale dependence explicitly.
In  addition to the Gouy phase $\chi_m^{|l|}$, the beam width varies continuously as $w(z)=w_0\sqrt{1+(z/z_R)^2}$  as it propagates along $z$;  $w_0$ is therefore the beam waist at the object plane $z=0$. (These and other properties of the classical LG beams have been reviewed thoroughly in the literature; see, Ref. \cite{book_twisted_photons_Padgett}.)

For completeness, the quantized field operators in the paraxial approximation, denoted with a  superscript $(0)$, for the magnetic Hertz potential, $\hat{\bm{Z}}_{1}^{(0)(+)}$ (cf. Eq. (\ref{eqn:Zs_bb})),  the vector field $\bm{A}_{\perp,1}^{(0)(+)}=\nabla\times \bm{Z}_1^{(0)(+)}$,  and the electric field $\bm{E}_{\perp, 1}^{(0)(+)}=-c^{-1}\partial_t\bm{A}_{\perp,1}^{(0)(+)}$ are listed below.  They are derived by  substituting  the paraxial form of  Eq. (\ref{eqn:Z_1s_f}), namely,
\begin{equation}
\mathcal{Z}_{1s,ml}(\bm{r},\omega)=f_\omega\tilde{\mathcal{Z}}_{1s,ml}^{(0)}=(f_\omega w_0)\mathcal{L}_{1s,ml}
\end{equation}
into Eq. (\ref{eqn:Zs_TG_1}), followed by taking the appropriate derivatives to get (note that the factor $f_\omega w_0=c/\omega$)
%
\begin{subequations}
\label{eqn:sol_paraxial}
\begin{eqnarray}
\bm{Z}^{(0)(+)}_{1}(\bm{r},t)&=&\sum_s\sum_{m,l}\int_0^\infty d\omega\left(\frac{c}{\omega}\right)\mathcal{A}_\omega e^{-i\omega(t-s z/c)}\hat{b}_{1s,ml}(\omega)\left(0,1,0\right)\mathcal{L}_{1s,ml}(\bm{r},\omega)
\label{eqn:Z0_1s_paraxial}\\
\bm{A}_{\perp 1}^{(0)(+)}&=&\sum_s\sum_{m,l}\int_0^\infty d\omega\,\mathcal{A}_\omega e^{-i\omega(t-s z/c)}\hat{b}_{1s,ml}(\omega) \left(s,\, 0,\, \frac{ic}{\omega}\frac{\partial}{\partial x} \right)\mathcal{L}_{1s,ml}(\bm{r},\omega)
\label{eqn:A0_1s_paraxial}\\
\bm{E}_{\perp 1}^{(0)(+)}&=&i\sum_s\sum_{m,l}\int_0^\infty d\omega\,\mathcal{E}_\omega e^{-i\omega(t-s z/c)}\hat{b}_{1s,ml}(\omega) \left(s,\, 0,\, \frac{ic}{\omega}\frac{\partial}{\partial x}\right)\mathcal{L}_{1s,ml}(\bm{r},\omega)
\label{eqn:E0_1s_paraxial}
\end{eqnarray}
\end{subequations}
%
The prefactor $\mathcal{A}_\omega=\sqrt{c\hbar/\omega}$ and $\mathcal{E}_\omega=(\omega/c)\mathcal{A}_\omega=\sqrt{\hbar\omega/c}$; the extra $\sqrt{c}$ factor in the denominator in $\mathcal{E}_\omega$, compared to the standard definition (see Ref. \cite{book_cohen}), compensates for the change in  dimension of $\hat{b}(\omega)\sim \hat{a}(\bm{k})/\sqrt{c}$ when transforming from the momentum to the frequency representation (cf. Eq. (\ref{eqn:a_IAS})). In Eqs. (\ref{eqn:A0_1s_paraxial}) and (\ref{eqn:E0_1s_paraxial}), the slowly varying $z$ derivatives of $\mathcal{L}$ are ignored. Note that $\bm{Z}_1$ is globally polarized along $\hat{\bm{y}}$, while $\bm{A}_{\perp 1}$ and $\bm{E}_{\perp 1}$ are strongly plane-polarized along $\hat{\bm{x}}$ with a small longitudinal component along $\hat{\bm{z}}$ in the paraxial limit \cite{Lax_paraxial}.

\end{widetext}


%
\end{document}